\pdfoutput=1
\documentclass{emulateapj}

\newcommand{\Kepler}{{\it Kepler}}

\newcommand{\degrees}{$^{\circ}$}

  \usepackage{graphicx}
\usepackage{xcolor}
\usepackage{hyperref}
\usepackage{breakurl}
\usepackage{amsmath}
\usepackage{graphicx}
\usepackage{physics}

\newcommand{\be}{\begin{equation}}
\newcommand{\ee}{\end{equation}}

\usepackage{xcolor}
\usepackage{hyperref}
\usepackage{breakurl}
\usepackage{amsmath}
\usepackage{graphicx}
\usepackage{verbatim}
\usepackage{booktabs}

\shorttitle{Cool Stars --- Hot Jupiters}
\shortauthors{Becker et al.}

\begin{document}

%% LaTeX will automatically break titles if they run longer than
%% one line. However, you may use \\ to force a line break if
%% you desire.

%A population of wide, coplanar giant companions to hot Jupiters around cool stars
\title{Exterior Companions to Hot Jupiters 
orbiting Cool Stars are Coplanar}
\author{
% Figs 1,3-4, compiled system properties / posteriors, secular and n-body sims, final Monte Carlo simulation and posterior construction, wrote text
Juliette C. Becker\altaffilmark{1,$\dagger$},
% Came up with original idea, sample identification, planning calculations / plots
Andrew Vanderburg\altaffilmark{2,3,$\dagger$, $\star$},
% Fred: ideological guidance, came up with ideas for n-body sims, wrote/edited text, spin argument
Fred C. Adams\altaffilmark{1,4},
% Figure 2, checks on secular theory, compiled HJ properties 
Tali Khain\altaffilmark{4},
%Marta (fit posteriors for trend systems)
Marta Bryan\altaffilmark{5},
}

%% Use \author, \affil, and the \and command to format
%% author and affiliation information.
%% from AASTeX v4.0. You can use \email to mark an email address
%% anywhere in the paper, not just in the front matter.
%% As in the title, use \\ to force line breaks. 

%% Notice that each of these authors has alternate affiliations, which
%% are identified by the \altaffilmark after each name.  Specify alternate
%% affiliation information with \altaffiltext, with one command per each
%% affiliation.
\altaffiltext{1}{Astronomy Department, University of Michigan, Ann Arbor, MI 48109, USA}
\altaffiltext{2}{Department of Astronomy, The University of Texas at Austin, 2515 Speedway, Stop C1400, Austin, TX 78712}
\altaffiltext{3}{Harvard--Smithsonian Center for Astrophysics, 60 Garden St., Cambridge, MA 02138, USA}
\altaffiltext{4}{Physics Department, University of Michigan, Ann Arbor, MI 48109, USA}
\altaffiltext{5}{California Institute of Technology, Pasadena, CA 91126}
\altaffiltext{$\dagger$}{NSF Graduate Research Fellow}
\altaffiltext{$\star$}{NASA Sagan Fellow}

%% Mark off your abstract in the ``abstract'' environment. In the manuscript
%% style, abstract will output a Received/Accepted line after the
%% title and affiliation information. No date will appear since the author
%% does not have this information. The dates will be filled in by the
%% editorial office after submission.

\begin{abstract}

The existence of hot Jupiters has challenged theories of planetary formation since the first extrasolar planets were detected. Giant planets are generally believed to form far from their host stars, where volatile materials like water exist in their solid phase, making it easier for giant planet cores to accumulate. Several mechanisms have been proposed to explain how giant planets can migrate inward from their birth sites to short-period orbits. One such mechanism, called Kozai-Lidov migration, requires the presence of distant companions in orbits inclined by more than $\sim40$ degrees with respect to the plane of the hot Jupiter's orbit. The high occurrence rate of wide companions in hot Jupiter systems lends support to this theory for migration. However, the exact orbital inclinations of these detected planetary and stellar companions is not known, so it is not clear whether the mutual inclination of these companions is large enough for the Kozai-Lidov process to operate. This paper shows that in systems orbiting cool stars with convective outer layers, the orbits of most wide planetary companions to hot Jupiters must be well aligned with the orbits of the hot Jupiters and the spins of the host stars. For a variety of possible distributions for the inclination of the companion, the width of the distribution must be less than $\sim20$ degrees to recreate the observations with good fidelity. As a result, the companion orbits are likely well-aligned with those of the hot Jupiters, and the Kozai-Lidov mechanism does not enforce migration in these systems. 
\end{abstract}
%As such, we dynamically constrain the true masses of the companions to hot Jupiters (WASP-41 and HAT-P-13).  
%The sample of known hot Jupiters around stars cooler than 6250 Kelvin with wide companions 

\keywords{ keywords,  planets and satellites: gaseous planets}

\section{Introduction}

Hot Jupiters, or Jupiter-sized planets orbiting with periods of a few days and distances of about 2-5\% of an astronomical unit, are an intriguing class of exoplanets with no analog in our own Solar system. Although hot Jupiters are thought to account for only about 0.9 -- 1.5\% of the total population of planets \citep{2005PThPS.158...24M,2008PASP..120..531C,2011arXiv1109.2497M,2012ApJ...753..160W,2015ApJ...799..229W}, they are over-represented in our current population of discovered exoplanets due to their large masses, large radii, and short orbital periods, which make them easy to detect in both transit and radial velocity observations. More than 300 hot Jupiters have been discovered to date\footnote{As of 12 August 2017, the NASA Exoplanet Archive reports 315 known hot Jupiters. \url{https://exoplanetarchive.ipac.caltech.edu/cgi-bin/TblView/nph-tblView?app=ExoTbls&config=planets}}. 

Since the discovery of the first hot Jupiters, understanding their origins has been a challenge for planet formation theorists, who have proposed several different mechanisms for how these planets form and how the systems are assembled into the architectures we see today. One traditional model for giant planet formation, which has been highly successfully applied to the formation of giant planets in our own solar system, is called core accretion \citep{1982P&SS...30..755S}. In this model, a small core (likely composed of rocky and dense volatile materials) first forms in the proto-planetary disk, far enough away from its host star that dense volatile materials like water and/or methane are in solid, rather than gaseous, forms. Once a core has formed and become massive enough, it subsequently accretes a massive hydrogen/helium dominated envelope via runaway gas accretion, leaving planets roughly the size and mass of Jupiter in orbits similar to that of Jupiter -- far away from their host stars. 

In this traditional picture, in order for the newly formed giant planet to become a  {\em hot} Jupiter, it must then migrate inwards towards its host star, halting its migration at an orbital distance of about 0.05 AU. Theorists have identified several mechanisms by which hot Jupiters might migrate from an orbit at tens of AU into their present-day short-period orbits. One migration mechanism involves torques arising from tidal-disk interactions (``disk toques"), which could cause the hot Jupiters to slowly spiral inwards towards their host stars in the plane of the protoplanetary disk \citep[see][]{2002ApJ...565.1257T, 2012ARA&A..50..211K}.  Another mechanism relies upon dynamical interactions between planets to excite high eccentricities in the proto-hot Jupiters after the gas disk has dissipated, bringing the planets into orbits whose perihelia distances are close to the surface of the host star. Tidal interactions when the planet comes close to the host star then might dissipate orbital energy, causing the orbit to shrink and result in the short-period orbits seen in hot Jupiter systems. There are a couple of ways to excite these high eccentricities. Eccentricity can be excited via the Kozai-Lidov effect, which we call Kozai-Lidov migration \citep[and which requires an inclined exterior companion;][]{1962P&SS....9..719L, 1962AJ.....67..591K}. In some, more rare, cases, eccentricity can also be excited via low-inclination secular interactions, which we call co-planar high eccentricity migration \citep{2015ApJ...805...75P}. 

Another recently revived mechanism for hot Jupiter formation is \emph{in situ} formation: instead of runaway accretion occurring far away from the host star, where dense volatile materials are abundant in their solid forms, the super-Earth-sized cores of the hot Jupiters form past the ice line, and migrate in to their modern orbital radii simultaneously with other material in the disk. At this new orbital radius, the gas surface density would then be high enough for the core to experience runaway gas accretion at that location, forming a hot Jupiter \citep{2016ApJ...829..114B}. This scenario builds on the idea that there exists a nearly ubiquitous population of super-Earth-sized planets orbiting close to their host stars \citep[e.g.][]{2013ApJ...766...81F}, many of which have sufficient mass to undergo run-away accretion. %If these super-Earths begin rapidly accreting gas from the disk as it migrates inwards, they might accrete enough to become hot Jupiters themselves. 

These four distinct mechanisms for hot Jupiter system assembly (disk torques, coplanar high-eccentricity migration, Kozai-Lidov high eccentricity migration, and \emph{in situ} formation) have different observational outcomes. High eccentricity migration destabilizes the orbits of close-in companions and requires the presence of distant massive companions in hot Jupiter systems which originally helped excite those high eccentricities. If Kozai-Lidov migration is dominant, then these companions should have mutual inclinations with the hot Jupiters of $\gtrsim$ 40 \degrees. By contrast, disk migration will likely result in dynamically quiet systems with low mutual inclinations. \emph{In situ} formation initially produces a coplanar inner system, but subsequent secular interactions may eventually produce systems with either aligned \citep{2016ApJ...829..114B} or misaligned \citep{2016ApJ...829..114B,2017AJ....154...93S} close-in exterior companions, such as those seen in the WASP-47 system \citep{becker}. Such interactions would not change the natal stellar obliquity. 

A powerful way to understand the architectures and formation histories of hot Jupiter systems is through measurements of or constraints on the angles between the orbital angular momentum and the stellar spin axis. The difference between these angles is commonly called the stellar obliquity. There is a striking observed correlation between the photospheric temperature of the host star and the stellar obliquity. Observations of hot Jupiters \citep{winnobliquity, albrecht}, \citep[and more tentatively, smaller planets as well,][]{mazeh} have shown that the orbits of planetary systems around cool stars ($T_\ast<6200$ K) tend to be aligned with the spin of the host star, while the orbits of planets around hot stars ($T_\ast>6200$ K) tend to be misaligned with the stellar spin axis. The boundary between the populations of hot and cool stars is commonly taken at stellar mass $M_\ast$ = 1.3 M$_{\odot}$, or equivalently at surface temperature $T_\ast$ = 6200 K. This threshold is often called the ``Kraft break'' \citep{kraft, faststarslowstar}. This mass limit corresponds to stellar configurations where the convective envelope becomes thin, which provides some clues to the physical processes involved. 

Although the observed pattern of obliquities as a function of stellar surface temperature remains under study as additional stellar obliquity measurements are performed with methods such as Doppler tomography \citep[recent measurements include][]{2017AJ....153..211Z, 2017arXiv170801291J} or the Rossiter-McLaughlin technique \citep{2005ApJ...622.1118O}, the fact that hot Jupiters around cool stars tend to have orbits that are well-aligned with their host stars' spins axes appears to hold. However this alignment came about, it is difficult to produce it by random chance, and similarly difficult to reproduce it once it has been disturbed. This alignment could be primordial \citep[for example, magnetic fields can realign a young star with its disk; see][]{2015ApJ...811...82S}, or it could come about by re-alignment of stellar spin axes due to the planets' tidal influence \citep{hut1980, adamsbloch, albrecht}, a fairly slow process which takes hundreds of millions of years or more \citep{albrecht, 2012MNRAS.423..486L}. Therefore, in order for hot Jupiters to maintain their spin/orbit alignment, their obliquities cannot be perturbed or changed on timescales significantly shorter than this benchmark value. 

In this paper, we ask the question: ``What effect do distant perturbing bodies have on the alignment of hot Jupiter orbits and the spins of their host stars?'' Many distant companions, both planetary and stellar, to hot Jupiters have been found, and in fact are more frequent around hot Jupiter hosts than around typical stars \citep{2014ApJ...785..126K, 2015ApJ...800..138N, 2016ApJ...821...89B}. These companions also seem to have little effect on the hot Jupiters' spin orbit alignments \citep{2014ApJ...785..126K, 2015ApJ...800..138N, 2016ApJ...821...89B}. But if these distant companions have a strong enough gravitational influence on the hot Jupiters and have large mutual inclinations, they could in principle disturb the spin orbit alignment of the hot Jupiters away from what we observe in cool stars. By calculating the effect of the observed distant companions to hot Jupiters, we can place constraints on the mutual inclination between these companions and the well aligned hot Jupiters. 

Here, we {\em statistically} constrain the orbital inclinations of exterior long-period companions in hot Jupiter systems. We approach this problem by identifying a sample of hot Jupiters orbiting cool stars with known long-period companions and measured stellar obliquity and calculating the probability that each of these hot Jupiters will retain its low inclination as a function of the inclination of the distant perturbing companions using secular and N-body techniques. In Section \ref{sec:methods}, we describe our sample selection and analysis techniques. In Section \ref{sec:res}, we present the statistical results of our analysis and show that most companions of hot Jupiters around cool stars orbit near the plane of the hot Jupiters' orbits. In Section \ref{sec:conc}, we discuss the implications of this result on hot Jupiter formation and suggest avenues for future work.

\section{Methods}

\label{sec:methods}

\subsection{Sample selection}

We focus in this paper on transiting hot Jupiters with known companions detected via radial velocity observations. Since the hot Jupiters transit, it is often possible to measure components of the stellar obliquity via the Rossiter McLaughlin effect, a crucial ingredient in our calculations. Also, because the hot Jupiters transit, we know their orbital inclinations quite precisely to be nearly 90\degrees. Therefore, any constraint on the orbital inclination of the distant companion constrains the mutual orientation of the two planets' orbits.

The systems we consider in this work are those with the following properties:
\begin{itemize}
\item The host star is cool (with an effective temperature below Kraft break; $T_\ast<6200$ K)
\item The star hosts a hot Jupiter \citep[roughly Jupiter-mass planet with an orbital period between 0.8 and 6.3 days; as defined in][]{steffenconstraints}
\item There exists in the literature a measure of either the projected or true stellar obliquity (angle between the stellar spin axis and the planet's orbital angular momentum vector) for the host star. We do not require this obliquity to have any particular value or precision, but merely for a measurement to exist.
\item There is evidence in the literature that the system has an additional companion in the system; this companion may be a Jupiter-like planet or a brown dwarf
\end{itemize}

\begin{figure*}[htbp] %  figure placement: here, top, bottom, or page
   \centering
   \includegraphics[width=7.0in]{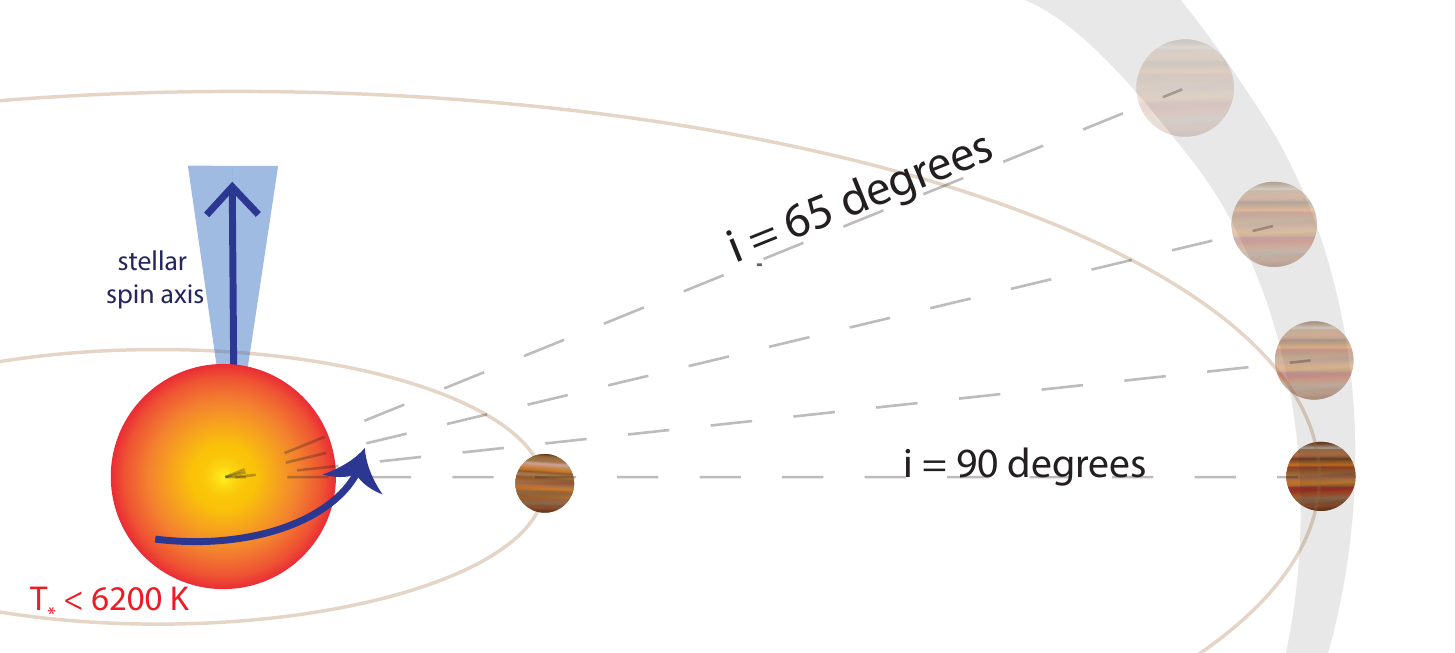} 
   \caption{A schematic diagram showing the orbital architecture of systems considered in this work. The systems we consider are those with stellar effective temperatures below the Kraft break ($T_\ast<6200 $K), a measured stellar obliquity, and evidence of an exterior companion whose residuals do not correlate with stellar activity level. The inclination of the outer companion is not known for any of the systems in our sample (this quantity is varied in the analysis).  }
   \label{fig:geometry}
\end{figure*} 

Figure \ref{fig:geometry} illustrates the geometry of the type of systems we consider in this work. A list of all the stars that fit these criteria and their properties, as well as the measured orbital properties of their planets, is given in Table \ref{bigtable}. Additional companions in these systems come in two forms: First, there are systems for which the orbits of additional companions have been well-characterized, and the period of their orbits are known (such as WASP-41, WASP-47, and HAT-P-13). Second, there are systems in which a trend in the RV data has been identified, but the (putative) companion does not have a precisely measured period (such as HAT-P-4, WASP-22, and WASP-53). These latter systems only have constraints on the companion's orbits \cite[see, for example, Figure 10 of][]{knutson14}, which can be derived from the radial velocity curves. In this work, we use the posteriors from \citet{2016ApJ...821...89B} for HAT-P-4 and WASP-22, and generate a new posterior for WASP-53 using the data in \citet{2017MNRAS.467.1714T} and the method from \citet{2016ApJ...821...89B}, without any adaptive optics constraints on outer companions. 

We exclude from our sample stars with companions and effective temperatures measured to be above the Kraft break. HAT-P-7, HAT-P-32, HAT-P-2 have temperatures right above Kraft break and have high projected obliquities, which is consistent with the convective realignment argument (the stars did not have a sufficient convective envelope to become realigned early in their lives). 
We exclude warm Jupiters (defined using the definition given in \citealt{steffenconstraints} to be Jupiter-mass planets with orbital periods between 6.3 days and 15.8 days) because these objects are not typically proposed to form through a high-eccentricity pathway and therefore, unlike hot Jupiters, are not expected to possess inclined companions \citep{2016ApJ...825...98H}.

The system XO-2N contains a hot Jupiter \citep{2007ApJ...671.2115B}, orbits a cool star, and has a projected stellar obliquity of 7 $\pm$ 11 degrees \citep{2015A&A...575A.111D}. \citet{knutson14} also presented RV evidence of a long-period signal in the system. However, \citet{2015A&A...575A.111D} found a correlation between the RV residuals and the stellar activity index $R_{HK}$, indicating that the long period signal is likely stellar activity and not a companion. For this reason, we also exclude this system from our sample (although we note that this system and its companion would fit perfectly into the aligned paradigm we see in our sample, were the companion to be real).

Of the systems we include, some have additional components that do not significantly effect the evolution of the system. The WASP-47 system is unique among hot Jupiter-hosting systems because it contains two short-period planets in addition to the hot Jupiter WASP-47b. Both of these additional planets are roughly coplanar with the hot Jupiter orbit (as they were both discovered via K2 transit photometry). In this work, we consider only the precession of the hot Jupiter, and do not impose additional constraints based on the transiting behavior or potential dynamical instability of the other planets \citep[unlike the analyses done in][]{beckeradams, vanderburg_masses}. We choose to consider the hot Jupiter alone because it is the planet for which the Rossiter-McLaughlin measurement was made \citep{sanchisojedaw47rm}. Excluding WASP-47 from the sample due to its unusual architecture would not change the results significantly since all hot Jupiters in our sample are aligned, so to maximize our sample size, we choose to include it. 

HAT-P-13 actually has three planets, a hot Jupiter and its two companions. The first companion has a period of 428.5 $\pm$ 3.0 days, an eccentricity of 0.691 $\pm$ 0.018, and an $m\sin{i}$ of 15.2 $\pm$ 0.1 $M_{J}$ \citep{2009ApJ...707..446B}. The second companion does not have a measured period, but an RV trend indicates its existence \citep{2010ApJ...718..575W}. The inner of those two (the middle body in the system) does not transit. Since the perturbations from the outermost body are expected to be adiabatic \citep{2013ApJ...778..100B}, we ignore the effect of the outer planet in our analysis. We do note that the influence of the outer planet has the potential to adiabatically misalign both inner planets. However, given that we measure a low stellar obliquity, and will show later that the middle planet is probably also aligned, it is unlikely the outer companion is highly inclined. Future modelling efforts may readily test this prediction.

\subsection{The Laplace-Lagrange Secular model}

\begin{figure}[htbp] %  figure placement: here, top, bottom, or page
   \centering
   \includegraphics[width=3.4in]{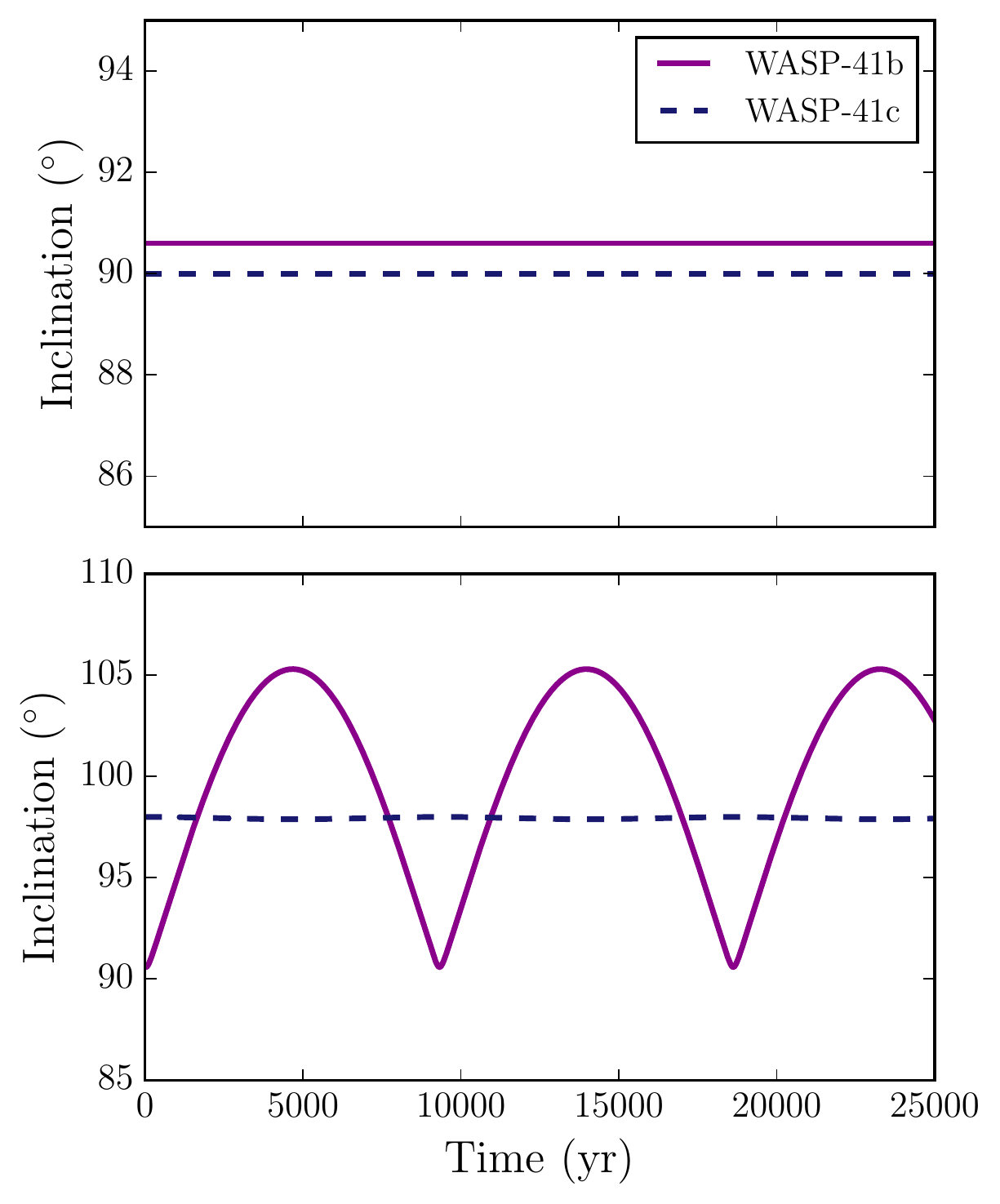} 
   \caption{The evolution of the inclination of WASP-41b and WASP-41c as given by Laplace-Lagrange secular theory: the secular theory was computed using inclinations centered around 0 degrees, and the inclination plotted is the secular result plus 90 degrees (to signify that WASP-41b is seen in an edge-on orbit). The companion's observed initial inclination differs in the two panels: $i_c = 90^{\circ}$ (top panel) and $i_c = 98^{\circ}$ (bottom panel). The presence of an inclined companion (-41c) results in an oscillating inclination angle for the hot Jupiter (-41b), affecting its angular momentum direction. In particular, a higher inclination of WASP-41c decreases the amount of time WASP-41b spends near its original orbital momentum direction, thereby increasing the likelihood of observing obliquity misalignments.}
   \label{fig:tali_figure}
\end{figure}

Additional exterior companions can alter the orbital inclination of the inner hot Jupiter through planet-companion interactions. As these interactions are mainly secular and non-resonant, we can approximate the system's orbital evolution over time using secular Laplace-Lagrange theory. This provides an approximation for the expected effect, which can be used to guide further analysis. Although we also use numerical N-body simulations (see below) to construct our final results in this work, this section outlines the analytic, guiding theory for describing the inclination evolution over time. 

As we expect secular interactions to dominate, we can construct a disturbing function for the planetary system, excluding terms that depend on the relative positions of the planets in their orbits \citep{1999ssd..book.....M}. The result is an equation which treats the planets as smeared-out rings of mass. Including only the terms describing the inclination of each planet's orbit to second order, this result becomes 
\be
\mathcal{R}^{\rm{(sec)}}_{j} = n_{j}a_{j}^{2} 
\Biggl[ \frac{1}{2} B_{jj} I_{j}^{2} 
+\sum_{k=1,\ j\ne k}^{N} \left( B_{jk} I_{j}I_{k} \cos{(\Omega_{j} -\Omega_{k})}\right)   
\Biggr]\,,
\ee
\noindent where $j$ is the planet number, $n$ the mean motion, $I$ the inclination, $\omega$ the argument of pericenter,
and $\Omega$ is the longitude of the ascending node. In the case of a spherical central body, the quantities $B_{ij}$ represent the interaction coefficients between pairs of planets and are given by 
\be
B_{jj} = -n_{j} \Biggl[ \frac{1}{4} \sum_{k=1,\ j\ne k}^{N} \frac{m_{k}}{M_{c} + m_{j}}
\alpha_{jk} \bar{\alpha}_{jk} b^{(1)}_{3/2}(\alpha_{jk}) \Biggr] \,,
\label{bmatrixdiag} 
\ee 
and 
\be
B_{jk} = n_{j} \left[ \frac{1}{4}\frac{m_{k}}{M_{c} + m_{j}}
  \alpha_{jk} \bar{\alpha}_{jk} b^{(1)}_{3/2}(\alpha_{jk}) \right]\,,
\label{bmatrixoff} 
\ee
where $m_{k}$ is the mass of the $k$th planet, $M_{c}$ is the mass of the central star, the  $\alpha_{jk}$ are the semi-major axis ratios $a_{j}/a_{k}$, and $\bar{\alpha}_{jk}$ are the semi-major axis ratios for $a_{j}/a_{k} < 1$. The function $b_{3/2}^{(1)}(\alpha)$ is 
the Laplace coefficient, which is defined by  
\be
b_{3/2}^{(1)}(\alpha)= \frac{1}{\pi} \int_{0}^{2\pi} 
\frac{\cos{\psi}\ d \psi}{(1-2 \alpha \cos{\psi} + \alpha^{2})^{3/2}}\,. 
\ee
Further explanation of this theory and potential expansions of the model can be found in \citet{1999ssd..book.....M}. 
Using the standard transformation
\be
p_j = I_j \sin \Omega_j \qquad {\rm and} \qquad 
q_j = I_j \sin \Omega_j \,,
\ee
the solutions of the eigenvalue problem defined by matrix \textbf{B} take the form:
\be
p_j (t) = \sum_{k=1}^N I_{jk} \sin (f_k t + \gamma_k) 
\label{ptime} 
\ee
and 
\be
q_j (t) = \sum_{k=1}^N I_{jk} \cos (f_k t + \gamma_k) \,. 
\label{qtime} 
\ee
To complete the initial condition problem, we define normalized
eigenvectors ${\cal I}_{jk}$ and corresponding scaling factors $T_k$ for the eigenvectors $I_{jk}$, 
\be
I_{jk} = T_k {\cal I}_{jk} \,, 
\ee
which allows us to use Equations (\ref{ptime}) and (\ref{qtime}) combined with the initial values of the inclination angles $I_j$ and the angles $\Omega_j$ for each planet to solve for the scaling factors $T_k$, i.e., 
\be 
p_j (t=0) = \sum_{k=1}^N T_k {\cal I}_{jk} \sin \gamma_k 
\ee
and
\be 
q_j (t=0) = \sum_{k=1}^N T_k {\cal I}_{jk} \cos \gamma_k \,.
\ee 
The result is an expression defining the time evolution of the
orbital inclination of each body in the system, 
\be 
I_{j}(t) = \sqrt{\left[ p_j(t) \right]^2 + 
\left[ q_j(t) \right]^2 } \,.
\label{inc_excite} 
\ee 

This equation can be used to generate the inclination evolution for planets in a system dominated by secular effects. By inspection, we see that the total angular momentum direction in the system will be conserved, but traded between planets in amounts mediated by the magnitude of the interaction coefficients. An example of the application of this theory is shown in Figure \ref{fig:tali_figure}, which plots the orbital inclination angles (as computed with the Laplace-Lagrange secular theory detailed above) over time for two realizations of WASP-41b and WASP-41c. The first system is considered to be co-planar, whereas the second case assumes that the companion WASP-41c is slightly inclined. We note that when Equation (\ref{inc_excite}) is used, the initial inclinations of transiting planets should be set to 0 degrees, rather than the 90 degrees traditionally reported observationally to denote edge-on orbits, due to the small angle approximation used in deriving the secular equations.

An inclined companion leads the orbit of hot Jupiter (-41b) to precess and allows the inclination angles to oscillate over time. A precessing hot Jupiter will appear aligned with its host's spin axis some (small) fraction of the time. This exact value depends on the observational error on the obliquity measurement as well as the orbital elements of all bodies in the system. As a result, for a single system, the fact that a hot Jupiter is aligned with the stellar spin axis does not completely specify the inclination of the companion. It is possible that our observations happen to occur at a moment in the secular cycles where the system passes through alignment. However, if we observe the entire population of hot Jupiter hosts to have spin axes aligned with their hot Jupiter's orbital angular momentum, then it is unlikely that their companions are highly inclined. In other words, the assessment of alignment in hot Jupiter systems must be done statistically.

\begin{figure}[htbp] 
   \centering
   \includegraphics[width=3.4in]{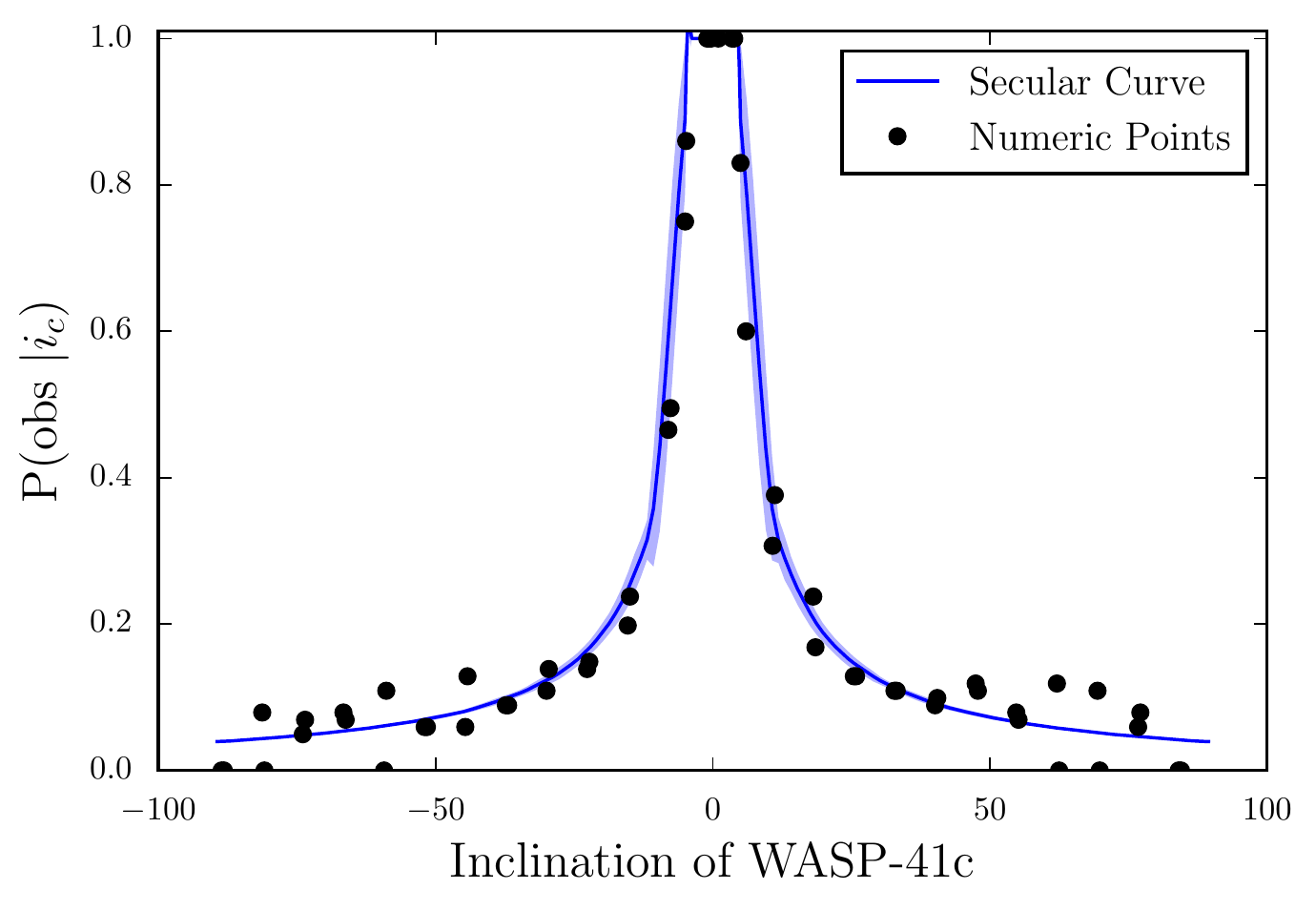} 
   \caption{Distribution for the probability of recreating the observations (obs), given some companion inclination for WASP-41c ($i_{c}$). This plots shows the comparison between the results computed from secular theory (solid blue curve) and the numerical N-body results (points). Here, an inclination of 0 degrees denotes an edge-on orbit (which observers report as having i=90 degrees). Except for the method used to generate the time series of orbital elements (secular theory versus the \texttt{Mercury6} N-body integrator), the probabilities in each case were generated the same way. The numerical simulations show good agreement with the secular calculation. The secular theory provides a robust motivation for this problem, and can be used to predict the interactions between planets at low mutual inclinations.}
   \label{fig:wasp41}
\end{figure}

\subsection{Numerical computation}
\label{sec:numcomp}
In addition to the secular theory described above, we ran numerical N-body integrations of these systems, as such simulations are capable of recovering orbital behavior that is not apparent from the secular theory alone. In Figure \ref{fig:wasp41}, we show the comparison between the results computed using each method for one system in our sample (WASP-41, the same system visualized in Figure \ref{fig:tali_figure}). The most important differences between the secular and numerical approaches are as follows: [1] The secular approach does not detect dynamical instabilities that result in ejections or collisions (see Figure \ref{fig:wasp41} - the points which lie on the x-axis are points where such a dynamical instability occurred, the inner planet was lost, and thus the system would never recreate observations). [2] The numerical approach allows for time-dependence in semi-major axis, while the secular theory does not. [3] The numerical approach will correctly capture the behavior of mean motion resonances should they arise (although, we expect these to be rare for the particular geometry of this problem). The differences between the secular and numerical results in Figure \ref{fig:wasp41} demonstrate that the secular theory is a good but not perfect approximation. To encompass all these behaviors, we treat the secular theory as motivation, and examine the behavior of each of the six systems in our sample using N-body integrations. In this numerical work, we use the system parameters and posteriors presented in Table \ref{bigtable}.

Another reason to use N-body methods rather than the secular approximation is that, although the numerical computations are very expensive, we only have six systems in our sample, and thus the calculation is feasible. Notice that, on average, a single trial using a \texttt{Mercury6} N-body simulation takes $10^{3}-10^{4}$ longer than the corresponding 
\texttt{python}-generated secular evolution. For longer integration times, this discrepancy grows larger.  Future analyses with a larger sample size might be able to use the secular approximation, which is generally accurate for sufficiently small mutual inclinations. 

The purpose of these numerical experiments is to examine the effect that varying the inclination of the companion has on the alignment of the spin axis of the star and the orbital angular momentum vector direction of the hot Jupiter. Recall that for a single system, we cannot draw firm conclusions about its orbital geometry from the fact the hot Jupiter transits today because precessing orbits allow planets to transit from a given line of sight with some duty cycle. Similarly, we must make an assumption about the underlying companion inclination distribution.
Since we are testing the population as a whole, and not just individual systems, we consider three possible priors for the population of companion inclinations: a Fisher distribution, a uniform distribution, and a delta function. 
For each distribution, we assign the width of the distribution to be $\sigma$, when $\sigma^{2} = \langle \sin^{2}{i} \rangle$, and the functional forms of each probability distribution $dp = f di$ and width are given as follows:
\begin{itemize}
    \item \textbf{Fisher distribution}. The Fisher distribution is often used \citep{2009ApJ...696.1230F, 2012AJ....143...94T} to describe the inclinations of planetary orbits, especially relative to the spin axis of their host star \citep[][see this paper for some illustrative plots describing the Fisher distribution]{2014ApJ...796...47M}. Its functional form can be written 
    \begin{equation}
    f_{f}(i | \kappa) = \frac{\kappa}{2 \sinh{\kappa}} e^{\kappa \cos{i}} \sin{i}\,,
    \label{probdis1}
    \end{equation}
    when $i$ is the orbital inclination angle. Then, we can find the width $\sigma$:
    \be
    \sigma^{2} = \langle \sin^{2}{i} \rangle = \int f_{\theta}(\theta | \kappa) \sin^{2}{i} di = 2 \frac{\coth{\kappa}}{\kappa} - \frac{2}{\kappa^{2}}
    \label{sigma2}
    \ee
    or
    \be
    \sigma =  \sqrt{2 \frac{\coth{\kappa}}{\kappa} - \frac{2}{\kappa^{2}}}
    \label{sigma1}
    \ee
    This form reduces to a Rayleigh distribution for large $\kappa$. For $\kappa \to 0$, the distribution becomes approximately isotropic.

    \item \textbf{Uniform distribution}. We assume that all companions come from a population with uniform scatter, but some maximum allowed inclination (defined as $\theta_{m}$). For each iteration, we generate companions by drawing from a uniform inclination distribution between a 0 degree mutual inclination and some maximum inclination. 
    The functional form for this distribution can be written as:
    \be
    \frac{dp}{di} = f_{u}(i | \theta_{m}) = \frac{1}{2\theta_{m}}
        \label{probdis2}
    \ee
    The width $\sigma$ of this distribution is again defined by the expectation value of $sin^{2}{i}$, where $i$ is the inclination drawn for each trial:
    \be
    \sigma^{2} = \langle \sin^{2}{i} \rangle = \int f_{u} \sin^{2}{i} di 
    \ee
    For a distribution ranging between $-\theta_{m}$ and $\theta_{m}$:
    \be
    \sigma^{2} = \int_{-\theta_{m}}^{\theta_{m}} \frac{1}{2\theta_{m}} \sin^{2}{i} di = \frac{\theta_{m} - \cos{\theta_{m}} \sin{\theta_{m}}}{2 \theta_{m}},
    \ee 
    or:
    \be
    \sigma = \sqrt{\frac{1}{2} - \frac{\sin{2\theta_{m}}}{4 \theta_{m}}}
    \label{sigma2b}
    \ee 
    \item \textbf{Delta function distribution}. We assume that all companions have the same inclination - so, the underlying companion distribution is a delta function at some inclination. This distribution has the probability function:
    \be
    \frac{dp}{di} = f_{\delta}(i | \theta_{x}) = \delta(i - \theta_{x})
        \label{probdis3}
    \ee
    and the width $\sigma$ can also be found:
    \be
    \sigma^{2} = \langle \sin^{2}{i} \rangle = \int f_{\delta} \sin^{2}{i} di = \sin^{2}{\theta_{x}}
    \ee
    So, the final width to the delta function companion distribution will be:
    \be
    \sigma = \sin{\theta_{x}}
    \label{sigma3}
    \ee
\end{itemize}

For each of those three priors, we initialized 1000 connected realizations of each of the six systems. (For example: a single realization includes all six planetary systems in independent integrations, all of which have inclinations drawn from the Fisher distribution of a given width. This process is then repeated 1000 times with different distributions widths. Then, the entire set of 1000 is repeated for each other prior type). In each realization, we sample from the known posteriors for each orbital element for all known planets. 
For the orbital elements of the hot Jupiter in each system, we set its initial inclination to be 90 degrees, and draw its orbital period, mass, eccentricity, and argument of periastron from observational priors (see Table \ref{bigtable}). For the orbital elements of the perturbing companions, we assign their orbital periods, masses, eccentricities, and arguments of periastron in the same way. We also draw an inclination for the perturbing companion from the prior being tested (either Equation [\ref{probdis1}], [\ref{probdis2}], or [\ref{probdis3}]). If a planet has an $m\sin{i}$ measurement instead of a true mass, we combine this measurement with our drawn inclination to find the true mass of the companion for that realization. 

After the initial conditions for the systems are specified, we evolve the systems using \texttt{Mercury6} \citep{m6} with time-steps set initially to be 1\% the orbital period of the innermost planet, and use the hybrid symplectic and Bulirsch-Stoer (B-S) integrator. We require energy conservation to a part in 10$^{-8}$ or better, and allow each integration to run for 10 Myr (which encompasses many secular periods). 

For each set of six systems, we then use the time-series of orbital elements computed by the N-body simulations to compute the projected stellar obliquity at each time-step. As each star in our sample has an observationally-measured projected obliquity, we then compute the probability that our simulated stellar obliquity would be measured to be consistent with this value (assuming the observational errors on our simulated measurement are equal to the error on the true measurement; see the third row of Table \ref{bigtable} for the projected obliquities and errors). The result of this computation is the probability that an observer would observe the stellar obliquity to be consistent with the true value we measure observationally at the current epoch.

\begin{figure*}[htbp] %  figure placement: here, top, bottom, or page
   \centering
   \includegraphics[width=7.0in]{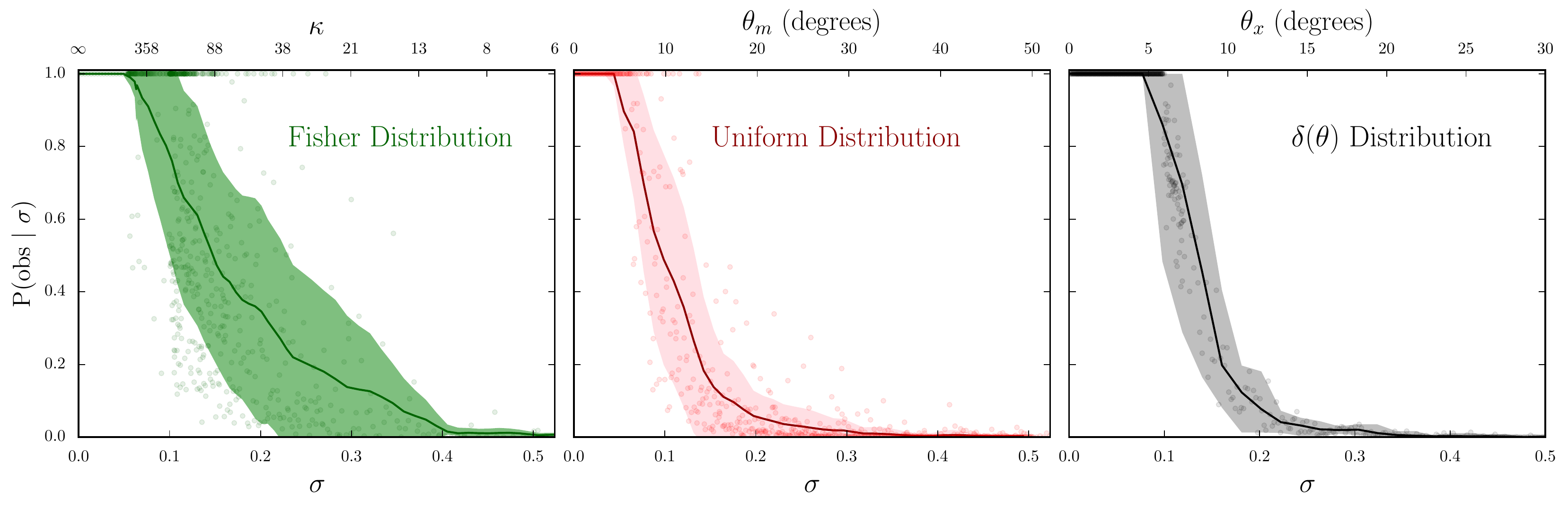} 
   \caption{For three different choices for the underlying companion inclination distributions with expectation value for orbital inclination of $\sigma^{2} = \langle \sin^{2}{i} \rangle$, these curves show the probability that we would measure the observed obliquities between the stellar spin axis and hot Jupiter orbital angular momentum for the entire population of systems in our sample (obs), given some companion inclination distribution $i_{c}$. Each panel uses a different prior on the type of distribution from which we draw the inclination of the companion orbit: the left panel is a Fisher distribution, the middle panel a uniform distribution with varying maximum inclinations, and the right panel is a delta function at each inclination. For all choices of the priors, the allowable range in inclination for the underlying distribution of the companion's orbit is less than $\sim20$ degrees out of the plane containing the hot Jupiter. }
   \label{fig:final_posterior}
\end{figure*}

Since each realization we have simulated includes six integrations (one for each planetary system), we then compute for each time-step the product of these six individual probabilities. This product is the probability that a simulated telescope making an observation at that time-step would measure a set of six projected stellar obliquities consistent with the true, current-day values. Then, using the entire time-series of probabilities, we compute a single marginalized probability $P(\rm{obs} | \sigma)$, when $\sigma$ is computed directly from the functional form of each prior (the final forms of which are given in Equation [\ref{sigma1}], Equation [\ref{sigma2}], and Equation [\ref{sigma3}]). This single probability describes the chance that we would observe all the same stellar obliquities presented in the third row of Table \ref{bigtable} given the prior we chose for the underlying distribution of companion inclinations.
We also plot in Figure \ref{fig:final_posterior} a smoothed curve representing the mean probability for each distribution width, with contoured error bars representing the 1$\sigma$ scatter at each distribution width.

\goodbreak

\section{Results}
\label{sec:res}

\subsection{The companion population tends to have nearly co-planar orbits} 

Figure \ref{fig:final_posterior} illustrates the main result of this work. We have considered systems containing hot Jupiters orbiting cool stars for which an obliquity measurement exists and which exhibit evidence for a companion. For three different types of distributions for the (unknown) inclination angle of the companion orbit, the numerical N-body simulations show that a large fraction of the cases with large mutual inclination angles result in a low probability of recreating the observations. 
As a result, it is unlikely that the companions to these cool hot Jupiter hosts generally have a high mutual inclination. Indeed, for all three prior choices (which range from the restrictive delta function distribution to the physically motivated and commonly used Fisher distribution), the allowable range of orbital planes for the companions is within 20--30 degrees of the orbital plane of the hot Jupiter: the probability curves in Figure \ref{fig:final_posterior} are plotted against $\sigma = \sqrt{\langle \sin^{2}{i} \rangle}$, and the top axis of each panel presents for physically intuitive units for each prior (the definitions of which can be found in Section \ref{sec:numcomp}). From these curves, we can compute the 95\% confidence interval for each prior, which will define an upper limit on the value we can expect $\sqrt{\langle \sin^{2}{i} \rangle}$ to assume, and then convert this to an angle, $i_{c}$, describing the likely maximum misalignment of exterior, coupled companions in these systems. For the Fisher distribution, this value is $i_{c}\sim$24 degrees. For the uniform distribution, this value is $i_{c}\sim$13 degrees. For the delta function distribution, this value is $i_{c}\sim$13 degrees. 

%\subsection{Inferences on the possible companion population}

Although the sample of known hot Jupiters with both stellar obliquity measurements and known exterior companions is small (only six such systems have been discovered at the time that this paper was written), we can nonetheless make significant inferences on the underlying distribution of possible orbital inclinations for the population of companions. The dynamical calculations carried out here show that, through primarily secular evolution, the inclination angles of the orbits are expected to evolve in the presence of an inclined companion. The fact that the stellar obliquity with respect to the hot Jupiter tends to be low constrains the secular evolution histories in these systems. If the underlying population of companions to systems containing hot Jupiters around cool stars has a random distribution of uniformly distributed inclination angles, the chance that our observations happened to catch the six known systems at times where the orbits of the hot Jupiters are aligned with the stellar spin axis is only $\sim10^{-7}$. As shown in Figure \ref{fig:final_posterior}, the orbits of the underlying companion population in these hot Jupiter systems are likely to be confined near the plane of the hot Jupiter orbit.   

\subsection{Implications for Hot Jupiter Formation and Migration}

Our conclusion that orbits of distant exterior companions to hot Jupiters are likely co-planar with hot Jupiter orbits has important implications for migration scenarios. The narrow distribution of inclination angles inferred here favors disk-driven migration mechanisms for hot Jupiters around cool stars. In this case, the disk and planets remain in nearly the same plane, and disk is generally aligned (within about 30 degrees) with the stellar spin axis (for additional discussion, see \citealt{2014ApJ...797...95L,becker,2016ApJ...825...98H,  2017AJ....153..265W} for discussions of alignment, and \citealt{2011MNRAS.412.2790L, 2014ApJ...797L..29S, 2014MNRAS.440.3532L,2015MNRAS.450.3306F} for mechanisms to excite misalignments, particularly with systems around hot stars). In situ formation of hot Jupiters would also lead to well-aligned planetary orbits \citep{2016ApJ...829..114B}. In contrast, high-eccentricity migration does not generally lead to low mutual inclinations. In this latter scenario, the migrating hot Jupiter attains high eccentricity, and hence a small periastron, so that tidal dissipation can circularize the orbit with a small semimajor axis. The mechanisms invoked to excite the high eccentricity --- including the Kozai-Lidov effect from stellar companions, planet-planet scattering, and secular interactions between planets --- generally result in high inclination configurations \citep{2007ApJ...669.1298F,2008ApJ...678..498N, 2011Natur.473..187N}. As these high-inclination configurations are found in hot Jupiter systems around hot stars ($T_\ast > 6200$ K), it is possible either that (a) hot stars, lacking a convective envelope, fail to realign the stellar spin-axis with orbital angular momentum early in their lives, or (b) the systems orbiting hot stars assemble via a different pathway.  
In either case, for cool stars, we favor a disk-driven migration scenario for dynamically coupled companions. 

On the other hand, some exceptions are possible \citep{2015ApJ...805...75P}, and the number of hot Jupiter systems for which we can carry out the analysis of this paper remains small. Fortunately, future observations should find an increasing number of hot Jupiter systems with additional companions orbiting cool stars. If these upcoming observations find a large number of misaligned systems, then high eccentricity migration will remain a viable alternative. On the other hand, if future observations find more systems with aligned obliquities, then it will support the paradigm advanced in this work of a coplanar companion population for cool hot Jupiter hosts.  

\subsection{Inclination of Companions to Hot Jupiters around Hot Stars}

In this paper, we only consider the inclinations of distant companions to hot Jupiters around cool stars because their obliquity angles are conveniently well aligned, making this type of analysis possible. This raises the question: ``Are companions to hot Jupiters around hot stars also coplanar?''

One possibility is that distant companions to hot Jupiters around hot stars are not well aligned with the hot Jupiters' orbits, and that their gravitational perturbations either cause or contribute to the the increased scatter in spin/orbit angles that are observed for these stars. This scenario hints at the explanation for correlation between stellar obliquity and stellar effective temperature by \citet{batyginmisalignment}, who suggests that the increased prevalence of stellar companions for more massive stars can explain the misalignment of hot Jupiter orbits with the spin axes of hot stars. \citet{batyginmisalignment} suggests that torques from distant, misaligned companions on the proto-planetary disks can cause the misalignments that are observed; our results demonstrate the well-known \citep[ex:][]{2017AJ....153...42L} result that closer misaligned companions can cause misalignments via secular interactions with the planet itself. 

Another possibility is that most distant companions to hot Jupiters around stars of all temperatures and masses are roughly co-planar with the hot Jupiters, and the large scatter in stellar obliquity observed in hot stars comes from some other mechanism. In this case, although the companions do not disturb the hot Jupiters' spin/orbit angles, we cannot tell because there is no apparent pattern for distant companions to disrupt. 

\subsection{Caveats}

The major caveats on the results quoted above can be summarized as follows. First, even in this paradigm, individual systems containing hot Jupiters around cool stars could (rarely) be found to have high-inclination companions due to unusual dynamical histories. For this reason, the methods and results of this work provide a statistical statement on the population of companions to hot Jupiters around cool stars, and cannot be used to determine true inclinations for individual systems.

Second, the temperature cut-off that we use in this work is chosen based on effective temperature. As these measurements are improved, some systems may move into or out of our sample. The ideal way to define the sample would be to include stars with thick convective envelopes, but currently, effective temperature is the best proxy for envelope size. As such, systems with host stars close to the temperature cut-off may be incorrectly categorized. 

Third, only dynamically coupled companions can be included in analyses of this nature. Companions with sufficiently large orbital radii may become decoupled from the dynamics of the inner system, and no longer affect the orbital precession of the hot Jupiter. Our statistical result does not apply to these very distant decoupled companions. Field surveys indicate that the occurrence rate of brown dwarfs (with masses ranging from 13 - 80 $M_J$) around Sun-like stars is low \citep{2014MNRAS.439.2781M}, with exact fractions ranging from 0.6\% to 0.8\%
\citep{2002ApJ...568..352V,2007ApJ...665..744P,2009AJ....137.3529W, 2011A&A...525A..95S}, suggesting that the companions for which we do not have fitted orbits (HAT-P-4, WASP-22, WASP-52) are more likely to be planetary companions rather than distant (potentially decoupled) brown dwarfs.

\section{Conclusion} 
\label{sec:conc}

In this work, we have shown statistically that distant exterior companions to hot Jupiters around cool stars must typically orbit in roughly the same plane as the hot Jupiter itself. Specifically, companion orbits must generally fall within 20 -- 30 degrees of the plane containing the hot Jupiter (see Figure \ref{fig:final_posterior})\footnote{Note that we expect any  additional planets to also be roughly in the same plane (see, for example, WASP-47).}. We constructed a sample of six hot Jupiter systems around cool stars (specifically, HAT-P-4, HAT-P-13, WASP-22, WASP-41, WASP-47, and WASP-53) and calculated the dynamical effects of distant perturbing companions as a function of the companion's orbital inclination. We performed a large ensemble of numerical simulations to show that if the inclination distribution companions to these systems extended much more than 20\degrees\ away from coplanar, then we would have been unlikely to observe the measured obliquities in our sample. We have also used secular theory for comparison; this approach is in good agreement with the full N-body simulations and can provide a time-saving alternative (see Figure \ref{fig:wasp41}). 

The fact that companions to hot Jupiters tend to orbit in nearly the same plane as the hot Jupiters themselves disfavors formation and migration models involving planet/planet scattering for hot Jupiters around cool stars. In particular, Kozai-Lidov migration typically requires a perturbing planetary (or stellar) companion with a mutual inclination of about 40\degrees\ or more. Mutual inclinations this large are strongly disfavored by our statistical analysis. This finding --- along with the fact that too few highly eccentric proto-hot Jupiters have been detected in \Kepler\ data to explain the hot Jupiter population \citep{dawson} --- suggests that Kozai-Lidov migration is not the dominant mode for forming hot Jupiter systems. Instead, this result favors formation scenarios that take place mostly within the plane of the proto-planetary disk, such as disk migration, {\em in situ} formation, or in some cases, coplanar high-eccentricity migration.

\acknowledgements

We thank Gongjie Li, Heather Knutson, Josh Winn, Ben Montet, Danielle Piskorz, Sarah Millholland, Clara Eng, and Iryna Butsky for useful conversations. We also thank Michael Dieterle for visualization suggestions, and thank the referee, Chris Spalding, for prompt and useful feedback. J.C.B and A.V. are supported by the NSF Graduate Research Fellowship grant nos. DGE 1256260 and 1144152, respectively. This work was performed in part under contract with the California Institute of Technology (Caltech)/Jet Propulsion Laboratory (JPL) funded by NASA through the Sagan Fellowship Program executed by the NASA Exoplanet Science Institute. This work used both the Extreme Science and Engineering Discovery Environment (XSEDE; NSF grant number ACI-1053575) and resources provided by the Open Science Grid, which is supported by the National Science Foundation and the U.S. Department of Energy's Office of Science.

\clearpage

\clearpage
\begin{turnpage}
\begin{deluxetable*}{lcccccc}
\tablewidth{0pt}
\tablehead{
  \colhead{ } & 
  \colhead{HAT-P-4}     &
  \colhead{HAT-P-13} &
  \colhead{WASP-22}     &
  \colhead{WASP-41}   &
  \colhead{WASP-47} & 
  \colhead{WASP-53}     \\
}
\startdata	
\\
\multicolumn{2}{l}{\textbf{Stellar Properties}} & & & & & \\					
$M_{*}$ ($M_{\odot}$)	&	1.26 $\pm$ 0.10 (1)	&	1.22 $^{+0.05}_{-0.10}$ (5)	&	1.249 $^{+ 0.088 }_{- 0.17 }$ (7)	&	0.987 $\pm$ 0.047 (7)	&	1.00 $\pm$ 0.05 (13)	&	0.87 $\pm$ 0.08 (17)	\\
$R_{*}$ ($R_{\odot}$)	&	1.617 $^{+0.057}_{-0.05}$ (1)	&	1.281 $\pm 0.079$ (5)	&	1.255 $^{+ 0.035 }_{- 0.034 }$ (7)	&	0.886 $\pm$ 0.017 (7)	&	1.15 $\pm$ 0.04 (14)	&	0.96 $\pm$ 0.24 (17)	\\
$\lambda$ (deg)	&	 -4.9 $\pm$ 11.9 (2)	&	1.9 $\pm$ 8.5 (6)	&	22 $\pm$ 16 (8)	&	6 $\pm$ 11 (7)	&	0 $\pm$ 24 (15)	&	 -4 $\pm$ 12 (17)	\\
$T_\ast$(K)	&	5860 $\pm$ 80 (1)	&	5653 $\pm$ 90 (5)	&	6153 $\pm$ 50 (7)	&	5546 $\pm$ 33 (7)	&	5400 $\pm$ 100 (16)	&	4950 $\pm$ 60 (17)	\\
\\
\multicolumn{2}{l}{\textbf{Hot Jupiter Properties}} & & & & & 	\\
Mass ($M_{J}$)	&	0.68 $\pm$ (1)	&	0.853 $^{+0.029}_{-0.046}$ (5)	&	0.617 $^{+ 0.033 }_{- 0.022 }$ (7)	&	0.977 $\pm$ 0.037 (7)	&	1.12 $\pm$ 0.04 (13)	&	2.132 $^{+ 0.09 }_{- 0.09 }$ (17)	\\
Period (days)	&	3.056536 $\pm5.7\times 10^{-5}$ (1)	&	2.916260 $\pm 1.0 \times 10^{-5}$ (5)	&	3.5327313 $\pm 5.8 \times 10^{-5}$ (8)	&	3.0524 $\pm$ 10$^{-5}$ (10)	&	4.15912 $\pm$ 10$^{-5}$ (13)	&	3.3098443 $\pm 2\times 10^{-6}$ (17)	\\
$e_{b}$	&	0 (1,2)	&	0.0133 $\pm$ 0.0041 (5)	&	0.023 $\pm$ 0.012 (9)	&	$<$ 0.026 (11)	&	0.0028 $\pm$ 0.0028 (13)	&	$<$ 0.03 (17)	\\
$i_{b}$ (deg.)	&	88.76 $^{+0.89}_{-1.38}$ (1)	&	83.4 $\pm$ 0.6 (5)	&	89.2 $\pm$ 0.5 (8)	&	89.4 $^{+ 0.3 }_{- 0.3 }$ (11)	&	89.2 $^{0.5}_{0.7}$	&	87.08 $^{+ 0.16 }_{- 0.15 }$ (17)	\\
$\omega_{b}$ (deg.)	&	 - 	&	181 $\pm$ 45 (5)	&	27 $^{51}_{-78}$	&	 - 	&	51 $\pm$ 82	&	 - 	\\
\\
\multicolumn{2}{l}{\textbf{Companion Properties}}  & & & & & \\				
$m \sin(i)$ ($M_{J}$)	&	 (3,4) $^{1}$	&	14.28 $\pm$ 0.28 (6)	&	 (3,4) $^{1}$	&	3.2 $\pm$ 0.20 (11)	&	1.24 $\pm$ 0.22 (11)	&	(17)$^{1}$	\\
Period (days)	&	(3,4)	&	446.27 $\pm$ 0.22 (6)	&	(3,4)	&	421 $\pm$ 2 (11)	&	572 $\pm$ 7 (11)	&	(17)	\\
$e_{c}$	&	(3,4)	&	0.6616 $\pm$ 0.0054 (6)	&	(3,4)	&	0.294 $\pm$ 0.024 (11)	&	0.13 $\pm$ 0.10 (11)	&	(17)	\\
$\omega_{c}$ (deg.)	&	(3,4)	&	176.7 $\pm$ 0.5 (5)	&	(3,4)	&353$\pm$6(11)&144$\pm$53(11)&	(17)	\\
	\enddata
	\tablecomments{Orbital parameters used for the analysis in this work, with the literature sources for each measured value. $^{1}$: this companion does not have a fitted orbit, but a trend indicating the presence of a companion with an orbital period longer than our observational baseline. In this work, we sample the orbital elements of these companions from the posteriors provided in \citet{2016ApJ...821...89B}. $^{2}$: Using the methodology in \citet{knutson14} and \citet{2016ApJ...821...89B}, we create a posterior of the same style for WASP-53, using the radial velocity measurements from \citet{2017MNRAS.467.1714T}. References: (1) \citealt{2007ApJ...670L..41K} (2) \citealt{2011AJ....141...63W} (3) \citealt{knutson14} (4) \citealt{2015ApJ...814..148P} (5) \citealt{2009ApJ...707..446B} (6) \citealt{2010ApJ...718..575W} (7) \citealt{2016MNRAS.457.4205S} (8) \citealt{2011A&A...534A..16A} (9) \citealt{2010AJ....140.2007M} (10) \citealt{2011PASP..123..547M} (11) \citealt{marion} (12) Vanderburg et al. (in prep) (13) \citealt{2017AJ....153..265W} (14) \citealt{becker} (15) \citealt{roberto} (16) \citealt{hellier} (17) \citealt{2017MNRAS.467.1714T} }
\label{bigtable}
\end{deluxetable*}
\end{turnpage}

\bibliographystyle{apj}

\begin{thebibliography}

\bibitem[{{Adams} \& {Bloch}(2015)}]{adamsbloch}
{Adams}, F.~C., \& {Bloch}, A.~M. 2015, \mnras, 446, 3676

\bibitem[{{Albrecht} {et~al.}(2012){Albrecht}, {Winn}, {Johnson}, {Howard},
  {Marcy}, {Butler}, {Arriagada}, {Crane}, {Shectman}, {Thompson}, {Hirano},
  {Bakos}, \& {Hartman}}]{albrecht}
{Albrecht}, S., {Winn}, J.~N., {Johnson}, J.~A., {et~al.} 2012, \apj, 757, 18

\bibitem[{{Anderson} {et~al.}(2011){Anderson}, {Collier Cameron}, {Gillon},
  {Hellier}, {Jehin}, {Lendl}, {Queloz}, {Smalley}, {Triaud}, \&
  {Vanhuysse}}]{2011A&A...534A..16A}
{Anderson}, D.~R., {Collier Cameron}, A., {Gillon}, M., {et~al.} 2011, \aap,
  534, A16

\bibitem[{{Bakos} {et~al.}(2009){Bakos}, {Howard}, {Noyes}, {Hartman},
  {Torres}, {Kov{\'a}cs}, {Fischer}, {Latham}, {Johnson}, {Marcy}, {Sasselov},
  {Stefanik}, {Sip{\H o}cz}, {Kov{\'a}cs}, {Esquerdo}, {P{\'a}l},
  {L{\'a}z{\'a}r}, {Papp}, \& {S{\'a}ri}}]{2009ApJ...707..446B}
{Bakos}, G.~{\'A}., {Howard}, A.~W., {Noyes}, R.~W., {et~al.} 2009, \apj, 707,
  446

\bibitem[{{Batygin}(2012)}]{batyginmisalignment}
{Batygin}, K. 2012, \nat, 491, 418

\bibitem[{{Batygin} {et~al.}(2016){Batygin}, {Bodenheimer}, \&
  {Laughlin}}]{2016ApJ...829..114B}
{Batygin}, K., {Bodenheimer}, P.~H., \& {Laughlin}, G.~P. 2016, \apj, 829, 114

\bibitem[{{Becker} \& {Adams}(2017)}]{beckeradams}
{Becker}, J.~C., \& {Adams}, F.~C. 2017, \mnras, 468, 549

\bibitem[{{Becker} \& {Batygin}(2013)}]{2013ApJ...778..100B}
{Becker}, J.~C., \& {Batygin}, K. 2013, \apj, 778, 100

\bibitem[{{Becker} {et~al.}(2015){Becker}, {Vanderburg}, {Adams}, {Rappaport},
  \& {Schwengeler}}]{becker}
{Becker}, J.~C., {Vanderburg}, A., {Adams}, F.~C., {Rappaport}, S.~A., \&
  {Schwengeler}, H.~M. 2015, \apjl, 812, L18

\bibitem[{{Bryan} {et~al.}(2016){Bryan}, {Knutson}, {Howard}, {Ngo}, {Batygin},
  {Crepp}, {Fulton}, {Hinkley}, {Isaacson}, {Johnson}, {Marcy}, \&
  {Wright}}]{2016ApJ...821...89B}
{Bryan}, M.~L., {Knutson}, H.~A., {Howard}, A.~W., {et~al.} 2016, \apj, 821, 89

\bibitem[{{Burke} {et~al.}(2007){Burke}, {McCullough}, {Valenti},
  {Johns-Krull}, {Janes}, {Heasley}, {Summers}, {Stys}, {Bissinger}, {Fleenor},
  {Foote}, {Garc{\'{\i}}a-Melendo}, {Gary}, {Howell}, {Mallia}, {Masi},
  {Taylor}, \& {Vanmunster}}]{2007ApJ...671.2115B}
{Burke}, C.~J., {McCullough}, P.~R., {Valenti}, J.~A., {et~al.} 2007, \apj,
  671, 2115

\bibitem[{{Chambers}(1999)}]{m6}
{Chambers}, J.~E. 1999, \mnras, 304, 793

\bibitem[{{Cumming} {et~al.}(2008){Cumming}, {Butler}, {Marcy}, {Vogt},
  {Wright}, \& {Fischer}}]{2008PASP..120..531C}
{Cumming}, A., {Butler}, R.~P., {Marcy}, G.~W., {et~al.} 2008, \pasp, 120, 531

\bibitem[{{Damasso} {et~al.}(2015){Damasso}, {Biazzo}, {Bonomo}, {Desidera},
  {Lanza}, {Nascimbeni}, {Esposito}, {Scandariato}, {Sozzetti}, {Cosentino},
  {Gratton}, {Malavolta}, {Rainer}, {Gandolfi}, {Poretti}, {Zanmar Sanchez},
  {Ribas}, {Santos}, {Affer}, {Andreuzzi}, {Barbieri}, {Bedin}, {Benatti},
  {Bernagozzi}, {Bertolini}, {Bonavita}, {Borsa}, {Borsato}, {Boschin},
  {Calcidese}, {Carbognani}, {Cenadelli}, {Christille}, {Claudi}, {Covino},
  {Cunial}, {Giacobbe}, {Granata}, {Harutyunyan}, {Lattanzi}, {Leto},
  {Libralato}, {Lodato}, {Lorenzi}, {Mancini}, {Martinez Fiorenzano},
  {Marzari}, {Masiero}, {Micela}, {Molinari}, {Molinaro}, {Munari}, {Murabito},
  {Pagano}, {Pedani}, {Piotto}, {Rosenberg}, {Silvotti}, \&
  {Southworth}}]{2015A&A...575A.111D}
{Damasso}, M., {Biazzo}, K., {Bonomo}, A.~S., {et~al.} 2015, \aap, 575, A111

\bibitem[{{Dawson} {et~al.}(2015){Dawson}, {Murray-Clay}, \&
  {Johnson}}]{dawson}
{Dawson}, R.~I., {Murray-Clay}, R.~A., \& {Johnson}, J.~A. 2015, \apj, 798, 66

\bibitem[{{Fabrycky} \& {Tremaine}(2007)}]{2007ApJ...669.1298F}
{Fabrycky}, D., \& {Tremaine}, S. 2007, \apj, 669, 1298

\bibitem[{{Fabrycky} \& {Winn}(2009)}]{2009ApJ...696.1230F}
{Fabrycky}, D.~C., \& {Winn}, J.~N. 2009, \apj, 696, 1230

\bibitem[{{Fielding} {et~al.}(2015){Fielding}, {McKee}, {Socrates},
  {Cunningham}, \& {Klein}}]{2015MNRAS.450.3306F}
{Fielding}, D.~B., {McKee}, C.~F., {Socrates}, A., {Cunningham}, A.~J., \&
  {Klein}, R.~I. 2015, \mnras, 450, 3306

\bibitem[{{Fressin} {et~al.}(2013){Fressin}, {Torres}, {Charbonneau}, {Bryson},
  {Christiansen}, {Dressing}, {Jenkins}, {Walkowicz}, \&
  {Batalha}}]{2013ApJ...766...81F}
{Fressin}, F., {Torres}, G., {Charbonneau}, D., {et~al.} 2013, \apj, 766, 81

\bibitem[{{Hellier} {et~al.}(2012){Hellier}, {Anderson}, {Collier Cameron},
  {Doyle}, {Fumel}, {Gillon}, {Jehin}, {Lendl}, {Maxted}, {Pepe}, {Pollacco},
  {Queloz}, {S{\'e}gransan}, {Smalley}, {Smith}, {Southworth}, {Triaud},
  {Udry}, \& {West}}]{hellier}
{Hellier}, C., {Anderson}, D.~R., {Collier Cameron}, A., {et~al.} 2012, \mnras,
  426, 739

\bibitem[{{Huang} {et~al.}(2016){Huang}, {Wu}, \&
  {Triaud}}]{2016ApJ...825...98H}
{Huang}, C., {Wu}, Y., \& {Triaud}, A.~H.~M.~J. 2016, \apj, 825, 98

\bibitem[{{Hut}(1980)}]{hut1980}
{Hut}, P. 1980, \aap, 92, 167

\bibitem[{{Johnson} {et~al.}(2017){Johnson}, {Cochran}, {Addison}, {Tinney}, \&
  {Wright}}]{2017arXiv170801291J}
{Johnson}, M.~C., {Cochran}, W.~D., {Addison}, B.~C., {Tinney}, C.~G., \&
  {Wright}, D.~J. 2017, ArXiv e-prints, arXiv:1708.01291

\bibitem[{{Kley} \& {Nelson}(2012)}]{2012ARA&A..50..211K}
{Kley}, W., \& {Nelson}, R.~P. 2012, \araa, 50, 211

\bibitem[{{Knutson} {et~al.}(2014{\natexlab{a}}){Knutson}, {Fulton}, {Montet},
  {Kao}, {Ngo}, {Howard}, {Crepp}, {Hinkley}, {Bakos}, {Batygin}, {Johnson},
  {Morton}, \& {Muirhead}}]{2014ApJ...785..126K}
{Knutson}, H.~A., {Fulton}, B.~J., {Montet}, B.~T., {et~al.}
  2014{\natexlab{a}}, \apj, 785, 126

\bibitem[{{Knutson} {et~al.}(2014{\natexlab{b}}){Knutson}, {Fulton}, {Montet},
  {Kao}, {Ngo}, {Howard}, {Crepp}, {Hinkley}, {Bakos}, {Batygin}, {Johnson},
  {Morton}, \& {Muirhead}}]{knutson14}
---. 2014{\natexlab{b}}, \apj, 785, 126

\bibitem[{{Kov{\'a}cs} {et~al.}(2007){Kov{\'a}cs}, {Bakos}, {Torres},
  {Sozzetti}, {Latham}, {Noyes}, {Butler}, {Marcy}, {Fischer}, {Fern{\'a}ndez},
  {Esquerdo}, {Sasselov}, {Stefanik}, {P{\'a}l}, {L{\'a}z{\'a}r}, {Papp}, \&
  {S{\'a}ri}}]{2007ApJ...670L..41K}
{Kov{\'a}cs}, G., {Bakos}, G.~{\'A}., {Torres}, G., {et~al.} 2007, \apjl, 670,
  L41

\bibitem[{{Kozai}(1962)}]{1962AJ.....67..591K}
{Kozai}, Y. 1962, \aj, 67, 591

\bibitem[{{Kraft}(1967)}]{kraft}
{Kraft}, R.~P. 1967, \apj, 150, 551

\bibitem[{{Lai}(2012)}]{2012MNRAS.423..486L}
{Lai}, D. 2012, \mnras, 423, 486

\bibitem[{{Lai}(2014)}]{2014MNRAS.440.3532L}
---. 2014, \mnras, 440, 3532

\bibitem[{{Lai} {et~al.}(2011){Lai}, {Foucart}, \& {Lin}}]{2011MNRAS.412.2790L}
{Lai}, D., {Foucart}, F., \& {Lin}, D.~N.~C. 2011, \mnras, 412, 2790

\bibitem[{{Lai} \& {Pu}(2017)}]{2017AJ....153...42L}
{Lai}, D., \& {Pu}, B. 2017, \aj, 153, 42

\bibitem[{{Lee} {et~al.}(2014){Lee}, {Chiang}, \&
  {Ormel}}]{2014ApJ...797...95L}
{Lee}, E.~J., {Chiang}, E., \& {Ormel}, C.~W. 2014, \apj, 797, 95

\bibitem[{{Lidov}(1962)}]{1962P&SS....9..719L}
{Lidov}, M.~L. 1962, \planss, 9, 719

\bibitem[{{Ma} \& {Ge}(2014)}]{2014MNRAS.439.2781M}
{Ma}, B., \& {Ge}, J. 2014, \mnras, 439, 2781

\bibitem[{{Marcy} {et~al.}(2005){Marcy}, {Butler}, {Fischer}, {Vogt}, {Wright},
  {Tinney}, \& {Jones}}]{2005PThPS.158...24M}
{Marcy}, G., {Butler}, R.~P., {Fischer}, D., {et~al.} 2005, Progress of
  Theoretical Physics Supplement, 158, 24

\bibitem[{{Maxted} {et~al.}(2010){Maxted}, {Anderson}, {Gillon}, {Hellier},
  {Queloz}, {Smalley}, {Triaud}, {West}, {Wilson}, {Bentley}, {Cegla}, {Collier
  Cameron}, {Enoch}, {Hebb}, {Horne}, {Irwin}, {Lister}, {Mayor}, {Parley},
  {Pepe}, {Pollacco}, {Segransan}, {Udry}, \& {Wheatley}}]{2010AJ....140.2007M}
{Maxted}, P.~F.~L., {Anderson}, D.~R., {Gillon}, M., {et~al.} 2010, \aj, 140,
  2007

\bibitem[{{Maxted} {et~al.}(2011){Maxted}, {Anderson}, {Collier Cameron},
  {Hellier}, {Queloz}, {Smalley}, {Street}, {Triaud}, {West}, {Gillon},
  {Lister}, {Pepe}, {Pollacco}, {S{\'e}gransan}, {Smith}, \&
  {Udry}}]{2011PASP..123..547M}
{Maxted}, P.~F.~L., {Anderson}, D.~R., {Collier Cameron}, A., {et~al.} 2011,
  \pasp, 123, 547

\bibitem[{{Mayor} {et~al.}(2011){Mayor}, {Marmier}, {Lovis}, {Udry},
  {S{\'e}gransan}, {Pepe}, {Benz}, {Bertaux}, {Bouchy}, {Dumusque}, {Lo Curto},
  {Mordasini}, {Queloz}, \& {Santos}}]{2011arXiv1109.2497M}
{Mayor}, M., {Marmier}, M., {Lovis}, C., {et~al.} 2011, ArXiv e-prints,
  arXiv:1109.2497

\bibitem[{{Mazeh} {et~al.}(2015){Mazeh}, {Perets}, {McQuillan}, \&
  {Goldstein}}]{mazeh}
{Mazeh}, T., {Perets}, H.~B., {McQuillan}, A., \& {Goldstein}, E.~S. 2015,
  \apj, 801, 3

\bibitem[{{Morton} \& {Winn}(2014)}]{2014ApJ...796...47M}
{Morton}, T.~D., \& {Winn}, J.~N. 2014, \apj, 796, 47

\bibitem[{{Murray} \& {Dermott}(1999)}]{1999ssd..book.....M}
{Murray}, C.~D., \& {Dermott}, S.~F. 1999, {Solar system dynamics}

\bibitem[{{Nagasawa} {et~al.}(2008){Nagasawa}, {Ida}, \&
  {Bessho}}]{2008ApJ...678..498N}
{Nagasawa}, M., {Ida}, S., \& {Bessho}, T. 2008, \apj, 678, 498

\bibitem[{{Naoz} {et~al.}(2011){Naoz}, {Farr}, {Lithwick}, {Rasio}, \&
  {Teyssandier}}]{2011Natur.473..187N}
{Naoz}, S., {Farr}, W.~M., {Lithwick}, Y., {Rasio}, F.~A., \& {Teyssandier}, J.
  2011, \nat, 473, 187

\bibitem[{{Neveu-VanMalle} {et~al.}(2016){Neveu-VanMalle}, {Queloz},
  {Anderson}, {Brown}, {Collier Cameron}, {Delrez}, {D{\'{\i}}az}, {Gillon},
  {Hellier}, {Jehin}, {Lister}, {Pepe}, {Rojo}, {S{\'e}gransan}, {Triaud},
  {Turner}, \& {Udry}}]{marion}
{Neveu-VanMalle}, M., {Queloz}, D., {Anderson}, D.~R., {et~al.} 2016, \aap,
  586, A93

\bibitem[{{Ngo} {et~al.}(2015){Ngo}, {Knutson}, {Hinkley}, {Crepp}, {Bechter},
  {Batygin}, {Howard}, {Johnson}, {Morton}, \&
  {Muirhead}}]{2015ApJ...800..138N}
{Ngo}, H., {Knutson}, H.~A., {Hinkley}, S., {et~al.} 2015, \apj, 800, 138

\bibitem[{{Ohta} {et~al.}(2005){Ohta}, {Taruya}, \&
  {Suto}}]{2005ApJ...622.1118O}
{Ohta}, Y., {Taruya}, A., \& {Suto}, Y. 2005, \apj, 622, 1118

\bibitem[{{Patel} {et~al.}(2007){Patel}, {Vogt}, {Marcy}, {Johnson}, {Fischer},
  {Wright}, \& {Butler}}]{2007ApJ...665..744P}
{Patel}, S.~G., {Vogt}, S.~S., {Marcy}, G.~W., {et~al.} 2007, \apj, 665, 744

\bibitem[{{Petrovich}(2015)}]{2015ApJ...805...75P}
{Petrovich}, C. 2015, \apj, 805, 75

\bibitem[{{Piskorz} {et~al.}(2015){Piskorz}, {Knutson}, {Ngo}, {Muirhead},
  {Batygin}, {Crepp}, {Hinkley}, \& {Morton}}]{2015ApJ...814..148P}
{Piskorz}, D., {Knutson}, H.~A., {Ngo}, H., {et~al.} 2015, \apj, 814, 148

\bibitem[{{Sahlmann} {et~al.}(2011){Sahlmann}, {S{\'e}gransan}, {Queloz},
  {Udry}, {Santos}, {Marmier}, {Mayor}, {Naef}, {Pepe}, \&
  {Zucker}}]{2011A&A...525A..95S}
{Sahlmann}, J., {S{\'e}gransan}, D., {Queloz}, D., {et~al.} 2011, \aap, 525,
  A95

\bibitem[{{Sanchis-Ojeda} {et~al.}(2015{\natexlab{a}}){Sanchis-Ojeda}, {Winn},
  {Dai}, {Howard}, {Isaacson}, {Marcy}, {Petigura}, {Sinukoff}, {Weiss},
  {Albrecht}, {Hirano}, \& {Rogers}}]{sanchisojedaw47rm}
{Sanchis-Ojeda}, R., {Winn}, J.~N., {Dai}, F., {et~al.} 2015{\natexlab{a}},
  \apjl, 812, L11

\bibitem[{{Sanchis-Ojeda} {et~al.}(2015{\natexlab{b}}){Sanchis-Ojeda},
  {Rappaport}, {Pall{\'e}}, {Delrez}, {DeVore}, {Gandolfi}, {Fukui}, {Ribas},
  {Stassun}, {Albrecht}, {Dai}, {Gaidos}, {Gillon}, {Hirano}, {Holman},
  {Howard}, {Isaacson}, {Jehin}, {Kuzuhara}, {Mann}, {Marcy}, {Miles-P{\'a}ez},
  {Monta{\~n}{\'e}s-Rodr{\'{\i}}guez}, {Murgas}, {Narita}, {Nowak}, {Onitsuka},
  {Paegert}, {Van Eylen}, {Winn}, \& {Yu}}]{roberto}
{Sanchis-Ojeda}, R., {Rappaport}, S., {Pall{\'e}}, E., {et~al.}
  2015{\natexlab{b}}, ArXiv e-prints, arXiv:1504.04379

\bibitem[{{Southworth} {et~al.}(2016){Southworth}, {Tregloan-Reed}, {Andersen},
  {Calchi Novati}, {Ciceri}, {Colque}, {D'Ago}, {Dominik}, {Evans}, {Gu},
  {Herrera-Cordova}, {Hinse}, {J{\o}rgensen}, {Juncher}, {Kuffmeier},
  {Mancini}, {Peixinho}, {Popovas}, {Rabus}, {Skottfelt}, {Tronsgaard},
  {Unda-Sanzana}, {Wang}, {Wertz}, {Alsubai}, {Andersen}, {Bozza}, {Bramich},
  {Burgdorf}, {Damerdji}, {Diehl}, {Elyiv}, {Figuera Jaimes}, {Haugb{\o}lle},
  {Hundertmark}, {Kains}, {Kerins}, {Korhonen}, {Liebig}, {Mathiasen}, {Penny},
  {Rahvar}, {Scarpetta}, {Schmidt}, {Snodgrass}, {Starkey}, {Surdej}, {Vilela},
  {Essen}, \& {Wang}}]{2016MNRAS.457.4205S}
{Southworth}, J., {Tregloan-Reed}, J., {Andersen}, M.~I., {et~al.} 2016,
  \mnras, 457, 4205

\bibitem[{{Spalding} \& {Batygin}(2015)}]{2015ApJ...811...82S}
{Spalding}, C., \& {Batygin}, K. 2015, \apj, 811, 82

\bibitem[{{Spalding} \& {Batygin}(2017)}]{2017AJ....154...93S}
---. 2017, \aj, 154, 93

\bibitem[{{Spalding} {et~al.}(2014){Spalding}, {Batygin}, \&
  {Adams}}]{2014ApJ...797L..29S}
{Spalding}, C., {Batygin}, K., \& {Adams}, F.~C. 2014, \apjl, 797, L29

\bibitem[{{Steffen} {et~al.}(2012){Steffen}, {Ragozzine}, {Fabrycky}, {Carter},
  {Ford}, {Holman}, {Rowe}, {Welsh}, {Borucki}, {Boss}, {Ciardi}, \&
  {Quinn}}]{steffenconstraints}
{Steffen}, J.~H., {Ragozzine}, D., {Fabrycky}, D.~C., {et~al.} 2012,
  Proceedings of the National Academy of Science, 109, 7982

\bibitem[{{Stevenson}(1982)}]{1982P&SS...30..755S}
{Stevenson}, D.~J. 1982, \planss, 30, 755

\bibitem[{{Tanaka} {et~al.}(2002){Tanaka}, {Takeuchi}, \&
  {Ward}}]{2002ApJ...565.1257T}
{Tanaka}, H., {Takeuchi}, T., \& {Ward}, W.~R. 2002, \apj, 565, 1257

\bibitem[{{Tremaine} \& {Dong}(2012)}]{2012AJ....143...94T}
{Tremaine}, S., \& {Dong}, S. 2012, \aj, 143, 94

\bibitem[{{Triaud} {et~al.}(2017){Triaud}, {Neveu-VanMalle}, {Lendl},
  {Anderson}, {Collier Cameron}, {Delrez}, {Doyle}, {Gillon}, {Hellier},
  {Jehin}, {Maxted}, {S{\'e}gransan}, {Smalley}, {Queloz}, {Pollacco},
  {Southworth}, {Tregloan-Reed}, {Udry}, \& {West}}]{2017MNRAS.467.1714T}
{Triaud}, A.~H.~M.~J., {Neveu-VanMalle}, M., {Lendl}, M., {et~al.} 2017,
  \mnras, 467, 1714

\bibitem[{{van Saders} \& {Pinsonneault}(2013)}]{faststarslowstar}
{van Saders}, J.~L., \& {Pinsonneault}, M.~H. 2013, \apj, 776, 67

\bibitem[{{Vanderburg} {et~al.}(2017){Vanderburg}, {Becker}, {Buchhave},
  {Mortier}, {Lopez}, \& {et al.}}]{vanderburg_masses}
{Vanderburg}, A., {Becker}, J.~C., {Buchhave}, L.~A., {et~al.} 2017, preprint

\bibitem[{{Vogt} {et~al.}(2002){Vogt}, {Butler}, {Marcy}, {Fischer},
  {Pourbaix}, {Apps}, \& {Laughlin}}]{2002ApJ...568..352V}
{Vogt}, S.~S., {Butler}, R.~P., {Marcy}, G.~W., {et~al.} 2002, \apj, 568, 352

\bibitem[{{Wang} {et~al.}(2015){Wang}, {Fischer}, {Horch}, \&
  {Huang}}]{2015ApJ...799..229W}
{Wang}, J., {Fischer}, D.~A., {Horch}, E.~P., \& {Huang}, X. 2015, \apj, 799,
  229

\bibitem[{{Weiss} {et~al.}(2017){Weiss}, {Deck}, {Sinukoff}, {Petigura},
  {Agol}, {Lee}, {Becker}, {Howard}, {Isaacson}, {Crossfield}, {Fulton},
  {Hirsch}, \& {Benneke}}]{2017AJ....153..265W}
{Weiss}, L.~M., {Deck}, K.~M., {Sinukoff}, E., {et~al.} 2017, \aj, 153, 265

\bibitem[{{Winn} {et~al.}(2010{\natexlab{a}}){Winn}, {Fabrycky}, {Albrecht}, \&
  {Johnson}}]{winnobliquity}
{Winn}, J.~N., {Fabrycky}, D., {Albrecht}, S., \& {Johnson}, J.~A.
  2010{\natexlab{a}}, \apjl, 718, L145

\bibitem[{{Winn} {et~al.}(2010{\natexlab{b}}){Winn}, {Johnson}, {Howard},
  {Marcy}, {Bakos}, {Hartman}, {Torres}, {Albrecht}, \&
  {Narita}}]{2010ApJ...718..575W}
{Winn}, J.~N., {Johnson}, J.~A., {Howard}, A.~W., {et~al.} 2010{\natexlab{b}},
  \apj, 718, 575

\bibitem[{{Winn} {et~al.}(2011){Winn}, {Howard}, {Johnson}, {Marcy},
  {Isaacson}, {Shporer}, {Bakos}, {Hartman}, {Holman}, {Albrecht}, {Crepp}, \&
  {Morton}}]{2011AJ....141...63W}
{Winn}, J.~N., {Howard}, A.~W., {Johnson}, J.~A., {et~al.} 2011, \aj, 141, 63

\bibitem[{{Wittenmyer} {et~al.}(2009){Wittenmyer}, {Endl}, {Cochran},
  {Ram{\'{\i}}rez}, {Reffert}, {MacQueen}, \& {Shetrone}}]{2009AJ....137.3529W}
{Wittenmyer}, R.~A., {Endl}, M., {Cochran}, W.~D., {et~al.} 2009, \aj, 137,
  3529

\bibitem[{{Wright} {et~al.}(2012){Wright}, {Marcy}, {Howard}, {Johnson},
  {Morton}, \& {Fischer}}]{2012ApJ...753..160W}
{Wright}, J.~T., {Marcy}, G.~W., {Howard}, A.~W., {et~al.} 2012, \apj, 753, 160

\bibitem[{{Zhou} {et~al.}(2017){Zhou}, {Bakos}, {Hartman}, {Latham}, {Torres},
  {Bhatti}, {Penev}, {Buchhave}, {Kov{\'a}cs}, {Bieryla}, {Quinn}, {Isaacson},
  {Fulton}, {Falco}, {Csubry}, {Everett}, {Szklenar}, {Esquerdo}, {Berlind},
  {Calkins}, {B{\'e}ky}, {Knox}, {Hinz}, {Horch}, {Hirsch}, {Howell}, {Noyes},
  {Marcy}, {de Val-Borro}, {L{\'a}z{\'a}r}, {Papp}, \&
  {S{\'a}ri}}]{2017AJ....153..211Z}
{Zhou}, G., {Bakos}, G.~{\'A}., {Hartman}, J.~D., {et~al.} 2017, \aj, 153, 211

\end{thebibliography}

\end{document}

%%
%% End of file `sample.tex'.